# Design of UAV flight state recognition and trajectory prediction system based on trajectory feature construction

Xingyu Zhou and Zhuoyong Shi

*Abstract*—With the impact of artificial intelligence on the traditional UAV industry, autonomous UAV flight has become a current hot research field. Based on the demand for research on critical technologies for autonomous flying UAVs, this paper addresses the field of flight state recognition and trajectory prediction of UAVs. This paper proposes a method to improve the accuracy of UAV trajectory prediction based on UAV flight state recognition and verifies it using two prediction models. Firstly, UAV flight data acquisition and data preprocessing are carried out; secondly, UAV flight trajectory features are extracted based on data fusion and a UAV flight state recognition model based on PCA-DAGSVM model is established; finally, two UAV flight trajectory prediction models are established and the trajectory prediction errors of the two prediction models are compared and analyzed after flight state recognition. The results show that: 1) the UAV flight state recognition model based on PCA-DAGSVM has good recognition effect. 2) compared with the traditional UAV trajectory prediction model, the prediction model based on flight state recognition can effectively reduce the prediction error.

*Index Terms*—UAV; Trajectory Prediction Model; PCA-DAGSVM Model; UAV Flight states Recognition Mode

## I. INTRODUCTION

UAV (unmanned aerial vehicle) is an aerial vehicle that does not require a pilot to be aboard [1, 2] and can be remotely controlled or autonomously pre-programmed to perform a variety of tasks. The development of UAVs has benefited from advances in a number of fields such as aviation technology, computer science, electronics and sensor technology. UAVs were initially used in the military as a tool for intelligence reconnaissance, targeting and aerial attack [3]. However, with the advancement of technology and the reduction of costs, drones are gradually being widely used in the civilian sector [4, 5]. Nowadays, UAVs play an important role in aerial photography, cargo transportation, agriculture, scientific research, disaster monitoring and rescue, and film and television filming [6, 7]. A safety monitoring system is a system that is used to supervise the parameters of the machine movement process, which allows the monitoring and control of the relevant parameters during the operation of the machine [8,9]. UAV trajectory prediction is the core technology of UAV safety supervision [10-12], which is necessary to achieve a more automated and accurate UAV supervision system. UAV trajectory prediction refers to the process of predicting the future trajectory of a UAV from the UAV's native information [13], which can provide more accurate navigation data for UAV control and navigation.

Flight state recognition of UAVs belongs to the field of pattern recognition, which was first widely used in the aerospace field to monitor the flight state of air and space electronic devices [14]. Related scholars in the field of pattern recognition use algorithms such as support vector machines [15], decision trees [16], random forests [17], and artificial neural networks [18] to conduct research.

In pattern recognition, Shi [19] combined pattern recognition technology with wearable devices and used an improved support vector machine model for pattern recognition of table tennis players' motor skills, and the results showed that the method was feasible in classifying motor skills. Rovinska [20] judged human emotional states based on physiological signals collected by wearable devices, and he used a self-encoder model to complete the recognition of human emotions, and the results show that the model improves the accuracy of recognition to a greater extent. Vellenga [21] develops a predictive forecasting model for 64 braking intentions of drivers during vehicle driving to address the challenges of achieving reliable driving intention recognition.

In trajectory prediction, Kan-nan [22] et al. proposed a low-complexity linear Kalman filter based on the differential state equation, improving trajectory prediction accuracy and response time. Wang [23] et al. developed a multi-step ballistic prediction method, MUKF, achieving real-time dynamic target prediction with an average prediction time of 20.16s. Zhang [24] et al. designed a four-dimensional trajectory prediction model based on historical flight data and UAV motion equations, incorporating a genetic algorithm for dynamic weighting Niu [25] et al. created an adjacent motion trajectory prediction algorithm based on model predictive control (MPC) without communication, surpassing traditional DMPC algorithms in simulation experiments. Xie [26] et al. proposed a Gaussian process regression (GPR)-based framework for online UAV trajectory prediction, outperforming other methods according to simulation and real data.

The above literature is a review of the current state of research by current scholars in the areas of pattern recognition and UAV trajectory prediction. Based on the need for research in areas related to autonomous flying UAVs, this paper investigates two key areas of autonomous flying UAVs, flight

Manuscript received Month xx, 2xxx; revised Month xx, xxxx; accepted Month x, xxxx. This work was supported in part by the … Department of xxx under Grant   (sponsor and financial support acknowledgment goes here).

Xingyu Zhou, and Zhuoyong Shi are with Northwestern Polytechnical University, Xi'an, China (e-mail: xingyuzhou@mail.nwpu.edu.cn; shizy@mail.nwpu.edu.cn; ).
 .

state recognition and UAV trajectory prediction.

The main contributions and innovations are listed below:

1 Designed the UAV record data acquisition system, which can complete the accurate monitoring of UAV data.

2 Improved the support vector machine model based on directed acyclic graph, and completed the identification of UAV flight state

3 The extraction of UAV trajectory features is completed by using data fusion.

4 Two trajectory prediction models corresponding to five different flight states are designed.

The rest of this article is organized as follows. The second part is the acquisition and pre-processing of UAV data. The third part is the UAV flight state recognition model. The fourth part is the UAV trajectory prediction model. The fifth section gives the prediction results of the two trajectory prediction models and explains them. Finally, Section 6 is conclusion.

## II. DATA ACQUISITION AND PROCESSING

### A. UAV data acquisition

Build UAV data acquisition system mainly includes position module air pressure module and attitude module, its structure is shown in Figure 1.

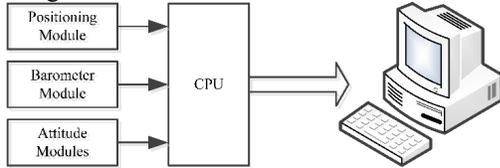

Fig. 1. UAV data acquisition system

The list of collected data is used for the study of this paper and the list of data is shown in Table 1.

TABLE I
ACQUISITION DATA OF UAV

| Data name | Data Volume | Time Resolution |
| --- | --- | --- |
| GPS data | 10000 | 0.1s |
| Barometer Data | 10000 | 0.1s |
| Acceleration Data | 10000 | 0.1s |
| Barometric Data | 10000 | 0.1s |

### B. Collection data pre-processing

Due to factors such as sensor acquisition error or data remote transmission process, the collected UAV trajectory data may generate noise, so it is necessary to carry out data preprocessing after collecting UAV data. Data preprocessing mainly includes three parts: eliminating abnormal data, repairing missing data and filtering high frequency noise, and finally outputting the repaired data after the preprocessing is completed.

1) Abnormal data rejection

For a set of data n samples $\{x_1, x_2, \ldots, x_n\}$ returned by the UAV, after returning the data, anomalous data needs to be culled from the retainer. According to the $3\sigma$ criterion, the data in the $(EX - 3\sigma, EX + 3\sigma)$ interval in $X_i$ is retained, the data outside the interval is considered to be abnormal data and eliminated, and the $X_i$ abnormal data outside the interval and the corresponding $x_{i+1}$ data are eliminated, so as to realize the elimination of abnormal data.

2) Missing data repair

After the outliers are removed, the data at the removed locations need to be repaired by interpolating the missing data using the Newton interpolation method, and finally the repaired data are output.

Based on Newton's interpolation method, a third-order difference quotient table is constructed to repair the function values of the missing data, so as to achieve data interpolation estimation of the missing data locations.

3) High frequency noise filtering

Based on that, an adaptive smoothing filtering method is used to update the filtering parameters by increasing the threshold value in order to adapt to the real-time signal. The new output value is defined as the current sample value and the previous output value weighted to obtain the current output value. The filtered output is shown in the formula.

$$Y(n) = m \bullet X(n) + (1-m) \bullet Y(n-1) \quad (1)$$

In formula(1), Y(n) is the filter output value, m is the filter coefficient between the interval [0,1], X(n) is the current sample value, and Y(n-1) is the previous filter output value.

## III. UAV FLIGHT STATES RECOGNITION MODEL

### A. Data fusion-based UAV trajectory feature extraction

Data fusion refers to the overall assessment that results from the detection and correlation of data information from multiple sources and levels and comprehensive analysis [27-33]. In the field of UAV aerial data, UAV data are collected by sensors as position information, angular velocity information, acceleration information, and air pressure information, and the motion parameters during UAV flight are fused and solved by data layer fusion so as to extract UAV trajectory characteristics. The UAV flight data fusion hierarchy is shown in Figure 2.

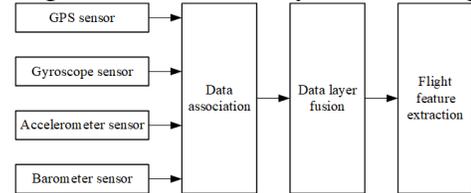

Fig. 2. UAV flight data fusion hierarchy

The motion parameters during UAV flight mainly include position in space, velocity, Euler angle, acceleration, acceleration direction, short-term curvature, etc.

The fusion of UAV flight data can achieve feature extraction of sensor data, and the extracted UAV flight data features include position, velocity, Euler angle, acceleration, direction of travel, and curvature in space.

The spatial position of the UAV is obtained by converting the barometric data to the z-axis position, and the spatial position of the UAV can be obtained by combining the plane position; the spatial velocity of the UAV is obtained by the first-order difference of the UAV position; the attitude angle of the UAV is calculated by the cumulative sum of the initial attitude and the angular velocity. The spatial acceleration of the UAV is obtained by the first-order difference of the velocity. The curvature of the drone is solved by the relationship between three consecutive sampling points. The definitions of position, velocity, Euler angle, acceleration, and curvature in UAV space and the calculation formulas are shown in Table 2.

TABLE II
DATA CHARACTERISTICS OF THE UAV DURING FLIGHT

| Data name | Definition | Calculation formula | |
|---|---|---|---|
| altitude | Position of UAV in space | $z = 4.43 \times 10^4 \times (1 - 9.87 \times 10^{-6} P)^{\frac{1}{5.256}}$ | (2) |
| speed | Change rate of UAV position with time | $\begin{cases} v_x(k) = \dfrac{\nabla x(k)}{T} \\ v_y(k) = \dfrac{\nabla y(k)}{T} \\ v_z(k) = \dfrac{\nabla z(k)}{T} \end{cases}$ | (3) |
| Posture angle | The angle of the UAV body coordinate system relative to the geodetic coordinate system | $\begin{cases} \theta_x(n) = \theta_x(0) + \sum_{i=1}^{n} \omega_x(i) \\ \theta_y(n) = \theta_y(0) + \sum_{i=1}^{n} \omega_y(i) \\ \theta_z(n) = \theta_y(0) + \sum_{i=1}^{n} \omega_y(i) \end{cases}$ | (4) |
| acceleration | Change rate of UAV speed with time | $\begin{cases} a_x(k) = \dfrac{\nabla v_x(k)}{T} = \dfrac{\nabla_2 x(k)}{T} \\ a_y(k) = \dfrac{\nabla v_y(k)}{T} = \dfrac{\nabla_2 y(k)}{T} \\ a_z(k) = \dfrac{\nabla v_z(k)}{T} = \dfrac{\nabla_2 z(k)}{T} \end{cases}$ | (5) |
| curvature | Curved degree of UAV motion curve | $k = \dfrac{2(x_1 y_2 + x_2 y_1 + x_3 y_1 - (y_1 x_2 + y_2 x_3 + y_3 x_1))}{abc}$ | (6) |

As shown in Table 2, the definitions of motion parameters such as position, velocity, Euler angle, acceleration and curvature in UAV space during UAV flight and their calculation formulas are described.

After filtering the UAV acquisition signal and extracting the data features to construct the UAV trajectory features, 75 feature quantities are extracted as the data features in Table 2. The mean, variance, maximum, minimum, peak and valley values of 13 data features such as position, velocity, Euler angle, acceleration and curvature in UAV space after adding the sliding window and the mean, variance, maximum, minimum, peak and valley values of synthetic velocity and synthetic acceleration are extracted respectively.

### B. UAV trajectory feature downscaling

In the field of machine learning, overlearning directly affects classification accuracy [34-36] and can increase the task size of machine learning. PCA (Principal Component Analysis) is a typical unsupervised dimensionality reduction algorithm that can reduce multiple metrics into several principal components [37, 38], this paper uses PCA method to reduce the dimensionality of the trajectory feature to construct an effective UAV trajectory feature.

Based on the metrics in the UAV flight data and the classification labels of the UAV flight states to form a matrix, each element of the matrix is subtracted from the mean value of that column of the matrix to obtain the decentered matrix as shown in formula(7).

$$X_{ij} = \begin{bmatrix} x_{11} - \dfrac{1}{m}\sum_{i=1}^{m} x_{i1} & x_{12} - \dfrac{1}{m}\sum_{i=1}^{m} x_{i2} & \cdots & x_{1n} - \dfrac{1}{m}\sum_{i=1}^{m} x_{in} \\ x_{21} - \dfrac{1}{m}\sum_{i=1}^{m} x_{i1} & x_{22} - \dfrac{1}{m}\sum_{i=1}^{m} x_{i2} & \cdots & x_{2n} - \dfrac{1}{m}\sum_{i=1}^{m} x_{in} \\ \vdots & \vdots & \ddots & \vdots \\ x_{m1} - \dfrac{1}{m}\sum_{i=1}^{m} x_{i1} & x_{m2} - \dfrac{1}{m}\sum_{i=1}^{m} x_{i2} & \cdots & x_{mn} - \dfrac{1}{m}\sum_{i=1}^{m} x_{in} \end{bmatrix} \quad (7)$$

The covariance matrix is calculated from the decentered matrix of the UAV flight data as shown in formula(8)

$$C = \dfrac{1}{m-1} X^T X \quad (8)$$

The eigenvalue $\lambda_k$ of the covariance matrix $C$ and the corresponding eigenvector $v_k$ are obtained by eigen decomposition of the covariance matrix $C$, which is shown in formula(9).

$$C v_k = \lambda_k v_k \quad (9)$$

The contribution of the i-th component to the total component $c_i$ is defined as shown in formula(10).

$$c_i = \dfrac{\lambda_i}{\sum_{k=1}^{p} \lambda_k} (i = 1, 2, \ldots, p) \quad (10)$$

In formula(10), $\lambda_i (i=1,2,\ldots,p)$ is the eigenvalue of the covariance matrix $R$.

The cumulative contribution margin is defined as shown in formula(11).

$$C_i = \dfrac{\sum_{k=1}^{i} \lambda_k}{\sum_{k=1}^{p} \lambda_k} (i = 1, 2, \ldots, p) \quad (11)$$

$c_i$ is sorted in order from largest to smallest, and the cumulative contribution is cumulated sequentially, and when

the cumulative contribution exceeds 85% of the cumulative contribution, it is used as the principal component of the motion feature constructed by the UAV.

Principal component analysis is used to construct UAV trajectory features to improve machine learning accuracy. The UAV trajectory features are analyzed and the information retention rate of the principal components is set to 85%, at which point the features are reduced to 14 dimensions.

*C. UAV flight states classification*

Flight states [39] are abstract descriptions of different flight behaviors during UAV flight, which can be reduced to several standard states such as climb, level flight, circling, turn and descent [40]. Classification is one of the core problems in data mining, machine learning and pattern recognition [41-44]. The identification of UAV flight states is a necessary preparation for the analysis of UAV operation, an aid for UAV maintenance and design optimization, and has an important practical value [45].

Support vector machine model is used to classify for the five flight states of UAVs, but the motion pattern of UAVs is not a binary case, so the traditional support vector machine model needs to be improved. Considering that there are only five flight states of UAVs, the indirect method is chosen to implement the multi-classification problem of UAV flight states by support vector machines. In the indirect method, the improved DAGSVM model based on directed acyclic graphs outperforms the one-to-one and one-to-many indirect classifiers in solving the score-based, omission-based classification problems. This paper classifies five flight states during UAV flight based on the DAGSVM (Directed Acyclic Graph Support Vector Machine) model. The five UAV flight states of climb, level flight, turn, circling and descent are recorded as A, B, C, D and E, respectively, and their corresponding directed acyclic diagrams are classified as shown in Figure 3.

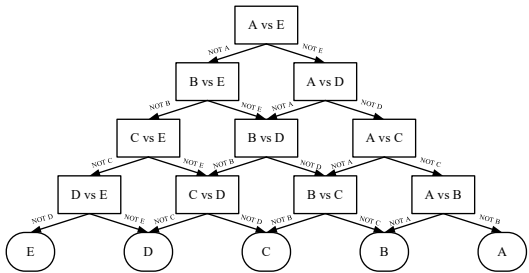

Fig. 3. Directed acyclic graph classification topology

As shown in Figure 3, the DAGSVM model-based directed acyclic graph structure is used to classify and identify the five flight states of the UAV 10 times to complete the identification of the UAV state.

*D. UAV flight states classification results*

Machine learning is performed by DAGSVM model to achieve the classification recognition of UAV flight states. The five motion modes of UAV include climb, level flight, circling, turn and descent. The input data are 5000 sets of 14-dimensional samples of UAV navigation data with known flight states labels during UAV navigation, including 1000 sets of each of the five flight states data, and the data are divided into training and test sets according to the ratio of 80% and 20%. The loss function uses a 0-1 loss function.

According to the Directed Acyclic Graph Support Vector Machine model in 3.3, the five flight states of climb, level flight, circling, turn and descent of the UAV are classified and identified. The confusion matrix heat map of the training and test sets of the classifier classification results is shown in Figure 4.

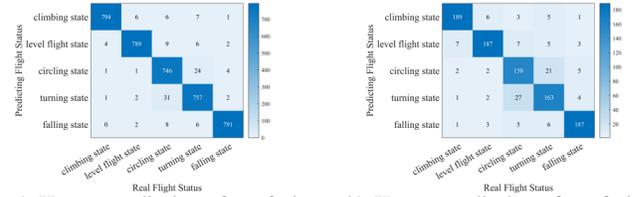

a) Heat map display of confusion matrix for training set    b) Heat map display of confusion matrix for testing set

Fig. 4. UAV flight states classification results

The F1 metric for any classification in this classifier is defined as the ratio of the true category of that classification to the linear combination of the predicted and true categories of that classification, and the F1 metric is shown in formula(12).

$$P = \frac{2TP}{2TP + 2\alpha FN + 2(1-\alpha)FP} \times 100\% \quad (12)$$

In equation (12), TP is the number of true categories in the type, FP is the number of predicted categories in the category but the true categories are not in the category, FN is the number of true categories in the category but the predicted categories are not in the category, and $\alpha$ is a weighting factor for precision and recall, taking values between 0,1, used to determine the preference for precision and recall in the final classifier's metric.

In UAV flight state recognition, it is more sensitive to the accuracy of true classification recognition, i.e., the weight for the accuracy rate should be greater than the recall rate, so set the weight coefficient $\alpha = 0.7$ for the accuracy rate and recall rate. The recognition results of the five flight states of climb, level flight, circling, turn and descent of the UAV are shown in Table 3.

TABLE III
TABLE OF CLASSIFICATION RESULTS FOR FIVE STATES

| Flight states | climbing | level flight | circling | turning | descenting |
|---|---|---|---|---|---|
| Training set accuracy rate | 97.54% | 97.41% | 96.13% | 95.46% | 98.02% |
| Training set recall | 99.25% | 98.63% | 93.25% | 94.63% | 98.88% |
| Training set F1 metric | 98.05% | 97.77% | 95.25% | 95.21% | 98.27% |
| Test Set Accuracy Rate | 92.65% | 89.47% | 84.04% | 82.74% | 92.57% |
| Test set recall | 94.50% | 93.50% | 79.00% | 81.50% | 93.50% |
| Test Set F1 Metric | 93.20% | 90.94% | 82.46 | 82.36 | 92.58% |

As shown in Figure 8 and Table 3 it can be seen that the DAGSVM classifier classifies well in climb, level flight and descent recognition, and slightly less well in circling motion states.

## IV. UAV TRAJECTORY PREDICTION MODEL

UAV trajectory prediction is the process by which a UAV predicts its own trajectory for a period of time in the future through relevant algorithms based on the already collected airborne information. UAV trajectory prediction can provide theoretical support and improve the accuracy of UAV control for UAV trajectory planning [46], UAV autonomous guidance [47], etc.

### A. Deep learning based UAV trajectory prediction model

The UAV operating state data is collected in different operating states as a neural network training data set. The trajectory prediction model established in this paper is used for trajectory prediction of small UAVs. The parameters of the prediction model and initial configuration are shown in Table 4.

TABLE IV
NEURAL NETWORK PARAMETER SETTING TABLE

| parameter name | number of settings |
|---|---|
| The number of training set samples | 80 |
| The number of samples in the test set | 20 |
| learning rate | 0.001 |
| Enter the number of nodes | 7 |
| Number of hidden layer nodes | 8 |
| Number of output nodes | 4 |
| activation function | ReLU |

### B. Multivariate Adams Predictive Correction Trajectory Prediction Model

During the flight of the UAV, there are many influences on the speed of the UAV in all directions, and in this model only the influence of time, spatial position and flight states on the speed is considered. Firstly, the UAV navigation speed is solved, then the differential equation of navigation speed about spatial position is established, and finally the UAV spatial position is solved by using Adams prediction correction formula. The multiple regression equations of the UAV flight speed about the trajectory position and the independent variable time are constructed respectively. The regression equation as shown in formula(13) is solved for the optimal solution with the error minimization as the mathematical model, which is shown in formula(13).

$$\min \|v(t,x) - V^*(t,x)\|_2^2 \quad (13)$$

In the best function approximation as in formula(13), the quadratic function approximation works best to construct the quadratic regression equation of velocity with respect to distance and time in the three directions of x y z as shown in formula(14).

$$\begin{cases} v_x = ax^2 + bt^2 + cxt + dx + et + f \\ v_y = ay^2 + bt^2 + cyt + dy + et + f \\ v_z = az^2 + bt^2 + czt + dz + et + f \end{cases} \quad (14)$$

The regression errors of the regression equations corresponding to the five flight states of the UAV are shown in Table 5.

TABLE 5
The five flight states of drones correspond to the error

| Flight states | Climbing | level flight | circling | turning | descenting |
|---|---|---|---|---|---|
| Errors/m | 0.0417m | 0.0594m | 0.0789 | 0.0441 | 0.0470 |

Based on the velocity binary function obtained by regression for each directional position, the Adams prediction correction formula is selected for iterative solution. The Adams prediction correction formula is shown in formula(15).

$$\begin{cases} y_{n+1}^{(0)} = y_n + \frac{h}{24}(55f_n - 59f_{n-1} + 37f_{n-2} - 9f_{n-3}) \\ y_{n+1} = y_n + \frac{h}{24}(9f(x_{n+1}, y_{n+1}^{(0)}) + 19f_n - 5f_{n-1} + f_{n-2}) \end{cases} \quad (15)$$

#### 1) Prediction algorithm design

The multivariate Adams prediction correction algorithm is shown in Algorithm 1

---
**Algorithm 1 Multivariate Adams Prediction Correction Algorithm**

input: training dataset $D = \{x_i, y_i\}_{i=1}^{N}$, validation dataset $V$

1 Determine the flight states of the UAV
2 Invoke the multivariate Adams prediction correction formula to the flight state
3 Determine initial velocity and position
4 repeat
5     Determine starting speed and position
4     for i=1…N do
5        Select sample $(t_i, x_i)$ from data set D
6
7        Predictive drone trajectory placement

$$y_{n+1}^{(0)} = y_n + \frac{h}{24}(55f_n - 59f_{n-1} + 37f_{n-2} - 9f_{n-3})$$

6        Correction of predicted trajectory

7
8        $y_{n+1} = y_n + \frac{h}{24}(9f(x_{n+1}, y_{n+1}^{(0)}) + 19f_n - 5f_{n-1} + f_{n-2})$

       Updated confidence interval
9     end
10 until Iterative update of all positions with speed

Out put $P = \{t_i, x_i\}_{i=N+1}^{M}$

---

#### 2) Confidence Curve Establishment

*a) Confidence curve radius determination*

The radius of the confidence interval of the speed regression equation for a single dimension of UAV navigation is calculated as shown in formula(16).

$$r_0 = S\sqrt{1 + \frac{1}{n} + \frac{(x_0 + \overline{x})^2}{\sum_{i=1}^{n}(x_i + \overline{x})^2}} \quad (16)$$

In formula(16), $S$ is the total variance, $n$ is the number of samples, $x_0$ is the specific variable, and $\overline{x}$ is the mean.

The confidence interval radius distance in space is the sum of the spatial distances of the confidence interval radius in each direction, and the spatial confidence interval radius is shown in formula(17).

$$r = \sqrt{r_x^2 + r_y^2 + r_z^2} \quad (17)$$

*b) Confidence curve direction determination*

The confidence curve for any point in space predicted at position $P_1(x_1, y_1, z_1)$ should satisfy orthogonal to the directions $P_0(x_0, y_0, z_0)$ and $P_1(x_1, y_1, z_1)$. That is, the unit normal vector of the face where this confidence curve is located can be expressed as shown in formula(18).

$$\begin{cases} a = \dfrac{x_1 - x_0}{\sqrt{(x_1-x_0)^2+(y_1-y_0)^2+(z_1-z_0)^2}} \\ b = \dfrac{y_1 - y_0}{\sqrt{(x_1-x_0)^2+(y_1-y_0)^2+(z_1-z_0)^2}} \\ c = \dfrac{z_1 - z_0}{\sqrt{(x_1-x_0)^2+(y_1-y_0)^2+(z_1-z_0)^2}} \end{cases} \quad (18)$$

As shown in Figure 9, $c_1$ is a confidence curve in space $P_0(x_0,y_0,z_0)$ and $P_1(x_1,y_1,z_1)$ are two points in space, $c_2$ is a curve with known parametric equations on the XoY plane past the origin, and the equation of the curve is shown in formula(19).

$$\begin{cases} x = r\cos(t) \\ y = r\sin(t) \\ z = 0 \end{cases} \quad (19)$$

The normal vectors $n_1$ and $n_2$ of the curves $c_1$ and $c_2$ have the relationship shown in formula(20).

$$G n_1^T = n_1^T \quad (20)$$

In formula(20), $G$ is the product of the Givens matrix, which can be expressed as shown in formula(21).

$$G = \begin{bmatrix} \cos\alpha & 0 & \sin\alpha \\ 0 & 1 & 0 \\ -\sin\alpha & 0 & \cos\alpha \end{bmatrix} \begin{bmatrix} \cos\beta & \sin\beta & 0 \\ -\sin\beta & \cos\beta & 0 \\ 0 & 0 & 1 \end{bmatrix} \quad (21)$$

In formula(21), $\alpha$ and $\beta$ are rotation factors, which can be expressed as shown in formula(22).

$$\begin{cases} \alpha = \arcsin(\dfrac{b}{\sqrt{a^2+b^2}}) \\ \beta = \arcsin(\dfrac{c}{\sqrt{a^2+b^2+c^2}}) \end{cases} \quad (22)$$

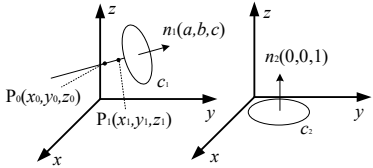

Fig. 5. Schematic diagram of confidence curve solution

By rotating the matrix $G$, the normal vector $n_1$ can be rotated to $n_2$. The inverse matrix $G^{-1}$ of this rotation matrix can be used to rotate the parameter equation of $c_2$ to the direction of $c_1$. The parameter equation in the direction of $c_1$ is shown in formula(23).

$$\begin{pmatrix} x \\ y \\ z \end{pmatrix} = G^{-1} \begin{pmatrix} r\cos(t) \\ r\sin(t) \\ 0 \end{pmatrix} \quad (23)$$

The direction of the confidence curve can be found by as shown in formula(23), but the spatial location of the confidence curve needs to be further determined.

*c) Confidence curve position determination*

The confidence curve can be rotated to the position where the center of the circle is at the origin by the rotation transformation as shown in formula (21), and the direction is the same as the direction of $P_1(x_1,y_1,z_1)$. Therefore it is necessary to shift the curve equation in formula (21) spatially, and the spatial shift is exactly the coordinates of $P_1(x_1,y_1,z_1)$, and the spatial shift can be expressed as shown in formula(24).

$$\begin{pmatrix} x \\ y \\ z \end{pmatrix} = G^{-1} \begin{pmatrix} r\cos(t) \\ r\sin(t) \\ 0 \end{pmatrix} + \begin{pmatrix} x_1 \\ y_1 \\ z_1 \end{pmatrix} \quad (24)$$

The confidence curve equation for any position in space can be expressed as in equation (24).

The method for solving the confidence curve is described in Algorithm 2.

| Algorithm 2 Confidence curve solving algorithm |
|---|
| Input: predicted location $P(x_i,y_i,z_i)_{i=1}^N$, confidence interval radius |
| 1  Determine the starting position of the UAV |
| 2  repeat |
| 3     Determine the predicted location |
| 4     for i=1…N do |
| 5        Calculate the spatial confidence curve radius |
| 6        Determine the direction-of-travel vector |
| 7        Determine the Givens matrix |
| 6        Solving for spatial displacement |
| 7        Determine the parametric equation of the curve |
| 8        Plot the confidence curve for point P |
| 9     end |
| 10 until Calculating the full confidence curve |
| Out put $c_{i=1}^N$ |

## V. UAV TRAJECTORY PREDICTION RESULTS AND ANALYSIS

### A. Neural Network Model Prediction Results

The flight data of the UAV is collected, preprocessed first, and then identified by PCA-SVM. The flight states of the UAV is predicted by climbing, leveling, turning, circling and descending flight states with a segment of the UAV navigation data, and the trajectory prediction results are shown in Figure 6.

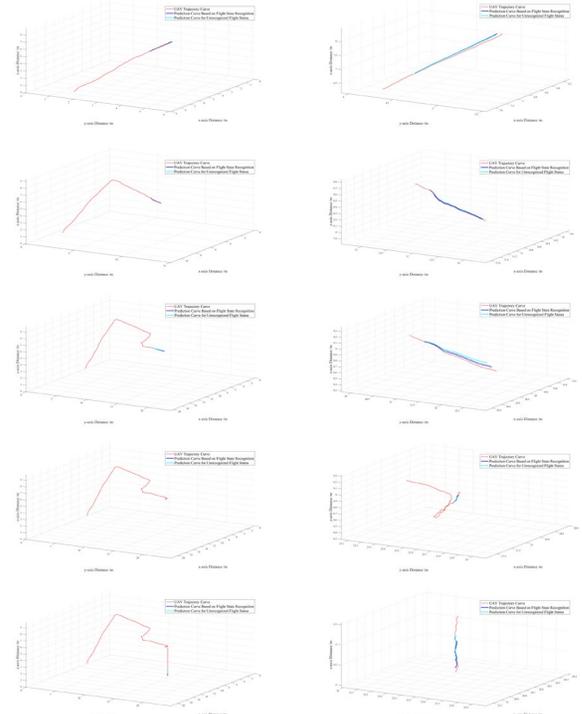

Fig. 6. Neural network model prediction results

### B. Multivariate Adams Predictive Prediction results

For the collected UAV flight data, after data preprocessing and flight states identification, the flight states are predicted in climb, level flight, turn, spiral and descent, respectively. The predicted and corrected flight trajectory prediction models are predicted by multivariate Adams in 4.2. The predicted model predicted and the predicted confidence curves of the UAV after flight states recognition are drawn, respectively, as shown in Figure 7.

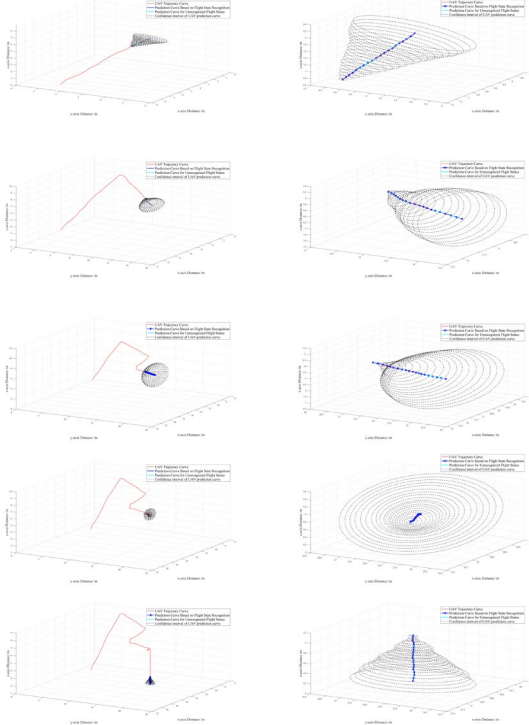

Fig. 7. Multivariate Adams model prediction results

### C. Analysis of UAV trajectory prediction results

The distance $d$ between the predicted trajectory point and the actual trajectory point is defined as shown in the formula(25).

$$d = \sqrt{(x-\hat{x})^2 + (y-\hat{y})^2 + (z-\hat{z})^2} \quad (25)$$

In the formula(25), $x$ represents the $x$-axis distance of the UAV's actual trajectory, $\hat{x}$ represents the $x$-axis distance of the UAV's predicted trajectory. $y$ represents the $y$-axis distance of the UAV's actual trajectory, and $\hat{y}$ represents the $y$-axis distance of the UAV's predicted trajectory. $z$ represents the $z$-axis distance of the UAV's actual trajectory, and $\hat{z}$ represents the $z$-axis distance of the UAV's predicted trajectory.

The average value of the Euclidean distance between the predicted trajectory and the actual trajectory at each time point is taken as the error distance of trajectory prediction. The error distance $\mu$ of the predicted trajectory is shown in the formula(26).

$$\mu = \frac{1}{N}\sum_{i=1}^{N} d_i = \sqrt{(x_i-\hat{x}_i)^2 + (y_i-\hat{y}_i)^2 + (z_i-\hat{z}_i)^2} \quad (26)$$

In the formula(26), N represents the number of trajectory points.

The predicted UAV trajectory errors before and after the flight states identification of the two prediction models in different flight states are calculated separately as shown in Figure 8.

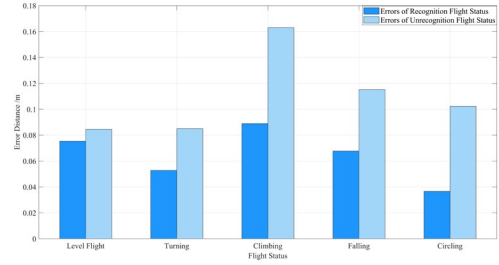

a) errors in neural network prediction models

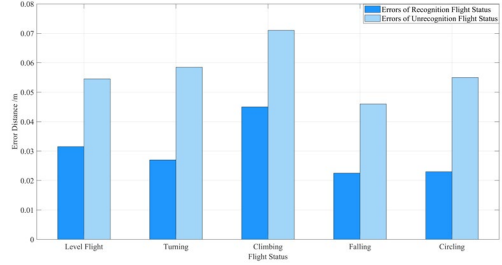

b) errors in Adams prediction models

Fig. 8. UAV trajectory prediction error diagram

As shown in Figure 8, the a) picture depicts the error of the neural network prediction model and the b) picture depicts the error of the Adams prediction model. Comparing the two images, it can be seen that: after the UAV flight state identification, the errors of both prediction models are significantly reduced. The test results show that the flight state recognition-based trajectory prediction model can achieve better UAV trajectory prediction than the direct adoption of the prediction model for UAV trajectory prediction.

## VI. CONCLUSION

In this paper, two prediction models are designed to predict UAV trajectories with and without flight states identification. The prediction results show that the prediction accuracy of the prediction model built with flight states identification is higher than that of the prediction model without flight states identification.

The specific work of this paper is as follows: firstly, data pre-processing, including abnormal data rejection, missing data interpolation and high-frequency noise filtering; secondly, a support vector machine model based on the improved support vector machine model of directed acyclic graph was established to identify the flight states of UAV; finally, five prediction models of flight states based on BP neural network and Adams prediction model were established to predict the trajectory respectively.

In future research, we will consider the use of unsupervised machine learning methods for UAV trajectory prediction, as well as trajectory prediction in multi-aircraft collaborative UAV flights.

## REFERENCES

[1] E. Kakaletsis et al., "Computer Vision for Autonomous UAV Flight Safety: An Overview and a Vision-based Safe Landing Pipeline Example," ACM Comput. Surv., vol. 54, no. 9, pp. 1–37, Dec. 2022, doi: 10.1145/3472288.

[2] C. Yan, L. Fu, X. Luo, and M. Chen, "A Brief Overview of Waveforms for UAV Air-to-Ground Communication Systems," in Proceedings of the 3rd International Conference on Vision, Image and Signal Processing, Vancouver BC Canada: ACM, Aug. 2019, pp. 1–7. doi: 10.1145/3387168.3387203.

[3] D. Orfanus, E. P. de Freitas, and F. Eliassen, "Self-Organization as a Supporting Paradigm for Military UAV Relay Networks," IEEE Commun. Lett., vol. 20, no. 4, pp. 804–807, Apr. 2016, doi: 10.1109/LCOMM.2016.2524405.

[4] W. W. Greenwood, J. P. Lynch, and D. Zekkos, "Applications of UAVs in Civil Infrastructure," J. Infrastruct. Syst., vol. 25, no. 2, p. 04019002, Jun. 2019, doi: 10.1061/(ASCE)IS.1943-555X.0000464.

[5] M. Sherman, M. Gammill, A. Raissi, and M. Hassanalian, "Solar UAV for the Inspection and Monitoring of Photovoltaic (PV) Systems in Solar Power Plants," in AIAA Scitech 2021 Forum, VIRTUAL EVENT: American Institute of Aeronautics and Astronautics, Jan. 2021. doi: 10.2514/6.2021-1683.

[6] A.-I. Siean, R.-D. Vatavu, and J. Vanderdonckt, "Taking That Perfect Aerial Photo: A Synopsis of Interactions for Drone-based Aerial Photography and Video," in ACM International Conference on Interactive Media Experiences, Virtual Event USA: ACM, Jun. 2021, pp. 275–279. doi: 10.1145/3452918.3465484.

[7] Y. Xia, G. Ye, S. Yan, Z. Feng, and F. Tian, "Application Research of Fast UAV Aerial Photography Object Detection and Recognition Based on Improved YOLOv3," J. Phys.: Conf. Ser., vol. 1550, no. 3, p. 032075, May 2020, doi: 10.1088/1742-6596/1550/3/032075.

[8] Y. Liu and S. Liu, "Design and Implementation of Farmland Environment Monitoring System Based on Micro Quadrotor UAV," J. Phys.: Conf. Ser., vol. 2281, no. 1, p. 012005, Jun. 2022, doi: 10.1088/1742-6596/2281/1/012005.

[9] M. Zhang, H. Wang, and J. Wu, "On UAV source seeking with complex dynamic characteristics and multiple constraints: A cooperative standoff monitoring mode," Aerospace Science and Technology, vol. 121, p. 107315, Feb. 2022, doi: 10.1016/j.ast.2021.107315.

[10] M. Corbetta, P. Banerjee, W. Okolo, G. Gorospe, and D. G. Luchinsky, "Real-time UAV Trajectory Prediction for Safety Monitoring in Low-Altitude Airspace," in AIAA Aviation 2019 Forum, Dallas, Texas: American Institute of Aeronautics and Astronautics, Jun. 2019. doi: 10.2514/6.2019-3514.

[11] P. Banerjee and M. Corbetta, "Uncertainty Quantification of Expected Time-of-Arrival in UAV Flight Trajectory," in AIAA AVIATION 2021 FORUM, VIRTUAL EVENT: American Institute of Aeronautics and Astronautics, Aug. 2021. doi: 10.2514/6.2021-2380.

[12] M. Zwick, M. Gerdts, and P. Stütz, "Sensor Model-Based Trajectory Optimization for UAVs Using Nonlinear Model Predictive Control," in AIAA SCITECH 2022 Forum, San Diego, CA & Virtual: American Institute of Aeronautics and Astronautics, Jan. 2022. doi: 10.2514/6.2022-1286.

[13] Z. Wang, G. Zhang, B. Hu, and X. Feng, "Real Time Detection and Identification of UAV Abnormal Trajectory," in Proceedings of the 2020 3rd International Conference on Artificial Intelligence and Pattern Recognition, Xiamen China: ACM, Jun. 2020, pp. 51–56. doi: 10.1145/3430199.3430212.

[14] P. Pietrzak, S. Szczęsny, D. Huderek, and Ł. Przyborowski, "Overview of Spiking Neural Network Learning Approaches and Their Computational Complexities," Sensors, vol. 23, no. 6, p. 3037, Mar. 2023, doi: 10.3390/s23063037.

[15] J. Zhang, Z. Shi, A. Zhang, Q. Yang, G. Shi, and Y. Wu, "UAV Trajectory Prediction Based on Flight State Recognition," IEEE Trans. Aerosp. Electron. Syst., pp. 1–16, 2023, doi: 10.1109/TAES.2023.3303854.

[16] A. Roy and S. Chakraborty, "Support vector machine in structural reliability analysis: A review," Reliability Engineering & System Safety, vol. 233, p. 109126, May 2023, doi: 10.1016/j.ress.2023.109126.

[17] P. G. Asteris et al., "Slope Stability Classification under Seismic Conditions Using Several Tree-Based Intelligent Techniques," Applied Sciences, vol. 12, no. 3, p. 1753, Feb. 2022, doi: 10.3390/app12031753.

[18] J. Hu and S. Szymczak, "A review on longitudinal data analysis with random forest," Briefings in Bioinformatics, vol. 24, no. 2, p. bbad002, Mar. 2023, doi: 10.1093/bib/bbad002.

[19] Z. Shi et al., "Design of motor skill recognition and hierarchical evaluation system for table tennis players," IEEE Sensors J., pp. 1–1, 2024, doi: 10.1109/JSEN.2023.3346880.

[20] S. Rovinska and N. Khan, "Affective State Recognition with Convolutional Autoencoders," in 2022 44th Annual International Conference of the IEEE Engineering in Medicine & Biology Society (EMBC), Glasgow, Scotland, United Kingdom: IEEE, Jul. 2022, pp. 4664–4667. doi: 10.1109/EMBC48229.2022.9871958.

[21] K. Vellenga, H. J. Steinhauer, A. Karlsson, G. Falkman, A. Rhodin, and A. C. Koppisetty, "Driver Intention Recognition: State-of-the-Art Review," IEEE Open J. Intell. Transp. Syst., vol. 3, pp. 602–616, 2022, doi: 10.1109/OJITS.2022.3197296.

[22] R. Kannan, "Orientation Estimation Based on LKF Using Differential State Equation," IEEE Sensors J., vol. 15, no. 11, pp. 6156–6163, Nov. 2015, doi: 10.1109/JSEN.2015.2455496.

[23] Y. Wang, K. Li, Y. Han, and X. Yan, "Distributed multi-UAV cooperation for dynamic target tracking optimized by an SAQPSO algorithm," ISA Transactions, vol. 129, pp. 230–242, Oct. 2022, doi: 10.1016/j.isatra.2021.12.014.

[24] H. Zhang, Y. Yan, S. Li, Y. Hu, and H. Liu, "UAV Behavior-Intention Estimation Method Based on 4-D Flight-Trajectory Prediction," Sustainability, vol. 13, no. 22, p. 12528, Nov. 2021, doi: 10.3390/su132212528.

[25] Z. Niu, X. Jia, and W. Yao, "Communication-Free MPC-Based Neighbors Trajectory Prediction for Distributed Multi-UAV Motion Planning," IEEE Access, vol. 10, pp. 13481–13489, 2022, doi: 10.1109/ACCESS.2022.3148145.

[26] G. Xie and X. Chen, "Efficient and Robust Online Trajectory Prediction for Non-Cooperative Unmanned Aerial Vehicles," Journal of Aerospace Information Systems, vol. 19, no. 2, pp. 143–153, Feb. 2022, doi: 10.2514/1.I010997.

[27] C. J. Okolie and J. L. Smit, "A systematic review and meta-analysis of Digital elevation model (DEM) fusion: pre-processing, methods and applications," ISPRS Journal of Photogrammetry and Remote Sensing, vol. 188, pp. 1–29, Jun. 2022, doi: 10.1016/j.isprsjprs.2022.03.016.

[28] T. Casian, B. Nagy, B. Kovács, D. L. Galata, E. Hirsch, and A. Farkas, "Challenges and Opportunities of Implementing Data Fusion in Process Analytical Technology—A Review," Molecules, vol. 27, no. 15, p. 4846, Jul. 2022, doi: 10.3390/molecules27154846.

[29] A. Tsanousa et al., "A Review of Multisensor Data Fusion Solutions in Smart Manufacturing: Systems and Trends," Sensors, vol. 22, no. 5, p. 1734, Feb. 2022, doi: 10.3390/s22051734.

[30] K. S. F. Azam, O. Ryabchykov, and T. Bocklitz, "A Review on Data Fusion of Multidimensional Medical and Biomedical Data," Molecules, vol. 27, no. 21, p. 7448, Nov. 2022, doi: 10.3390/molecules27217448.

[31] S. R. Stahlschmidt, B. Ulfenborg, and J. Synnergren, "Multimodal deep learning for biomedical data fusion: a review," Briefings in Bioinformatics, vol. 23, no. 2, p. bbab569, Mar. 2022, doi: 10.1093/bib/bbab569.

[32] Z. Wei, F. Zhang, S. Chang, Y. Liu, H. Wu, and Z. Feng, "MmWave Radar and Vision Fusion for Object Detection in Autonomous Driving: A Review," Sensors, vol. 22, no. 7, p. 2542, Mar. 2022, doi: 10.3390/s22072542.

[33] J. G. A. Barbedo, "Data Fusion in Agriculture: Resolving Ambiguities and Closing Data Gaps," Sensors, vol. 22, no. 6, p. 2285, Mar. 2022, doi: 10.3390/s22062285.

[34] W. Jia, M. Sun, J. Lian, and S. Hou, "Feature dimensionality reduction: a review," Complex Intell. Syst., vol. 8, no. 3, pp. 2663–2693, Jun. 2022, doi: 10.1007/s40747-021-00637-x.

[35] R. Rani, M. Khurana, A. Kumar, and N. Kumar, "Big data dimensionality reduction techniques in IoT: review, applications and open research challenges," Cluster Comput, vol. 25, no. 6, pp. 4027–4049, Dec. 2022, doi: 10.1007/s10586-022-03634-y.

[36] P. J. Schmid, "Dynamic Mode Decomposition and Its Variants," Annu. Rev. Fluid Mech., vol. 54, no. 1, pp. 225–254, Jan. 2022, doi: 10.1146/annurev-fluid-030121-015835.

[37] J. Moreira, B. Silva, H. Faria, R. Santos, and A. S. P. Sousa, "Systematic Review on the Applicability of Principal Component Analysis for the Study of Movement in the Older Adult Population," Sensors, vol. 23, no. 1, p. 205, Dec. 2022, doi: 10.3390/s23010205.

[38] V. Bruni, M. L. Cardinali, and D. Vitulano, "A Short Review on Minimum Description Length: An Application to Dimension Reduction

[38] in PCA," Entropy, vol. 24, no. 2, p. 269, Feb. 2022, doi: 10.3390/e24020269.
[39] H.-J. Jeong, S.-Y. Choi, S.-S. Jang, and Y.-G. Ha, "Probability machine-learning-based communication and operation optimization for cloud-based UAVs," J Supercomput, vol. 76, no. 10, pp. 8101–8117, Oct. 2020, doi: 10.1007/s11227-018-2728-4.
[40] A. Giagkos, E. Tuci, M. S. Wilson, and P. B. Charlesworth, "UAV flight coordination for communication networks: genetic algorithms versus game theory," Soft Comput, vol. 25, no. 14, pp. 9483–9503, Jul. 2021, doi: 10.1007/s00500-021-05863-6.
[41] M. Lochner and B. Bassett, "Machine Learning for Transient Classification: Workshop 13," Proc. IAU, vol. 14, no. S339, pp. 274–274, Nov. 2017, doi: 10.1017/S1743921318002740.
[42] Y. Song, Z. Hu, T. Li, and H. Fan, "Performance Evaluation Metrics and Approaches for Target Tracking: A Survey," Sensors, vol. 22, no. 3, p. 793, Jan. 2022, doi: 10.3390/s22030793.
[43] X. Niu, X. Yuan, Y. Zhou, and H. Fan, "UAV track planning based on evolution algorithm in embedded system," Microprocessors and Microsystems, vol. 75, p. 103068, Jun. 2020, doi: 10.1016/j.micpro.2020.103068.
[44] Q. Yang, Z. Ye, X. Li, D. Wei, S. Chen, and Z. Li, "Prediction of Flight Status of Logistics UAVs Based on an Information Entropy Radial Basis Function Neural Network," Sensors, vol. 21, no. 11, p. 3651, May 2021, doi: 10.3390/s21113651.
[45] R. K. Mehra, S. Seereeram, J. T. Wen, and D. S. Bayard, "Nonlinear predictive control for spacecraft trajectory guidance, navigation and control," in AIP Conference Proceedings, Albuquerque, New Mexico (USA): AIP, 1998, pp. 147–152. doi: 10.1063/1.54919.
[46] D. González-Arribas, M. Soler, and M. Sanjurjo-Rivo, "Robust Aircraft Trajectory Planning Under Wind Uncertainty Using Optimal Control," Journal of Guidance, Control, and Dynamics, vol. 41, no. 3, pp. 673–688, Mar. 2018, doi: 10.2514/1.G002928.
[47] N. Dadkhah and B. Mettler, "Survey of Motion Planning Literature in the Presence of Uncertainty: Considerations for UAV Guidance," J Intell Robot Syst, vol. 65, no. 1–4, pp. 233–246, Jan. 2012, doi: 10.1007/s10846-011-9642-9.

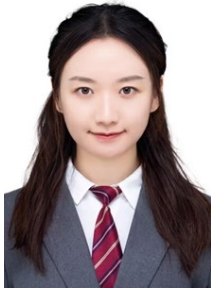
**Xingyu Zhou received the B.S. degree in Electronic Information Engineering from Xi'an University of Technology in 2020. She is currently pursuing the M.E. degree in the School of Electronics and Information Engineering at Northwestern Polytechnical University.**
**Her research interests include reinforcement learning, intelligent decision-making for unmanned aerial vehicles (UAVs), and Artificial Intelligence(AI).**

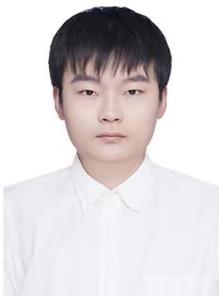
**Zhuoyong Shi received his bachelor's degree in electronic information engineering from Xi'an Jiaotong University City College. He is currently pursuing the master degree in electronic science and technology at Northwestern Polytechnical University, China.**

**His research interests include reinforcement learning, intelligent decision-making for unmanned aerial vehicles (UAVs), and Artificial Intelligence(AI).**